\documentclass[aps,prd,preprint,showpacs,preprintnumbers]{revtex4-1}

\usepackage[nointlimits,nosumlimits]{amsmath}
\usepackage{amsfonts,amssymb,amsthm,ifthen}
\usepackage[utf8x]{inputenc}

\newcommand\mN{\mathcal N}
\newcommand\mE{\mathcal E}
\newcommand\mV{\mathcal V}
\newcommand\mO{\mathcal O}
\newcommand\mQ{\mathcal Q}

\newcommand{\dd}[2][]{\frac{\partial #1}{\partial #2}}

\newcommand{\scal}[3][]{\ifthenelse{\equal{#1}{}}{
  \left\langle #2,\,#3 \right\rangle
}{\ifthenelse{\equal{#1}{(}}{
  \left( #2,\,#3 \right)
}{\ifthenelse{\equal{#1}{[}}{
  \left[ #2,\,#3 \right]
}{
  #1\left( #2,\,#3 \right)
}}}}

\newcommand{\arr}[2]{%
  \begin{array}{@{}#1@{}}#2\end{array}}

\newcommand{\abs}[1]{\left| #1 \right|}

\newcommand\tlambda{\tilde{\lambda}}
\newcommand\tpsi{\tilde{\psi}}
\newcommand\dbeta{\dot{\beta}}
\newcommand\ophi{\overline{\phi}}

\begin{document}

\preprint{HU-EP-12/19}
\title{Vertex Operators of Super Wilson Loops}
\author{Josua Groeger}
\affiliation{Humboldt-Universit\"at zu Berlin,\\
Institut f\"ur Mathematik and Institut f\"ur Physik\\
Rudower Chaussee 25, 12489 Berlin, Germany}
\email{groegerj@mathematik.hu-berlin.de}
\date{\today}

\begin{abstract}
We study the supersymmetric Wilson loop as introduced by Caron-Huot, which
attaches to lightlike polygons certain edge and vertex operators, whose
shape is determined by supersymmetry constraints.
We state explicit formulas for the vertex operators to all orders in the
Graßmann expansion, thus filling a gap in the literature.
This is achieved by deriving a recursion formula
out of the supersymmetry constraints.
\end{abstract}

\pacs{11.30.Pb, 11.25.Tq, 12.60.Jv}

\maketitle

\section{Introduction}

Gluon scattering amplitudes have been known to be dual to Wilson loops along lightlike polygons.
While first shown at strong coupling (Ref.~\onlinecite{AM07}) through the famous AdS/CFT duality
introduced in Ref.~\onlinecite{Mal98}, this result has later been verified at weak coupling
(Refs.~\onlinecite{DKS08,BHT08}). For a review, consult Ref.~\onlinecite{AR08}.

Recently, a similar duality (at weak coupling) between the full superamplitude of
$\mN=4$ super Yang-Mills theory and a supersymmetric extension of the Wilson loop
has been claimed, of which two variants appeared almost simultaneously.
The first approach by Mason and Skinner (Ref.~\onlinecite{MS10}) originates in momentum twistor
space and translates into the integral over a superconnection in spacetime.
The second is due to Caron-Huot (Ref.~\onlinecite{CH11}) and attaches
to lightlike polygons certain edge and vertex operators, whose
shape is determined by supersymmetry constraints.
At the classical level, both approaches
are identical only on-shell (Ref.~\onlinecite{BKS12}).
Belitsky showed in Ref.~\onlinecite{Bel12} that the conjectured duality with superamplitudes
indeed holds, however only upon subtracting an anomalous contribution
from the super Wilson loop.

The operators in the Caron-Huot approach depend on momentum supertwistors.
While the edge operators are wellknown, explicit formulas for the vertex operators
have been available in the literature only up to fourth order in the Graßmann expansion.
The aim of this article is to fill this gap. We state explicit formulas for the
vertex operators up to maximum order. This is achieved by deriving a recursion
formula out of the supersymmetry constraints.

To fix notation, we let $W_n$ denote the super Wilson loop and
$\mE_i$ and $\mV_{i,i+1}$ the edge and vertex operators, respectively,
which depend on the (odd) momentum supertwistors $\eta_i^A$ and $\eta_{i+1}^A$.
At zeroth order, the ordinary Wilson loop should be recovered, thus leading to the ansatz
$\mE_i=p_i\cdot A+\mO(\eta)$ and $\mV_{i,i+1}=1+\mO(\eta)$.
The supersymmetry constraints are such that $\mQ_A^{\alpha}W_n=0$ is to vanish,
where the $\mQ_A^{\alpha}=q_A^{\alpha}+c_0\sum_i\lambda_i^{\alpha}\dd{\eta_i^A}$ act
on the fields as well as the momentum supertwistors. This is achieved if the
edges and vertices transform by an infinitesimal super gauge transformation
\begin{subequations}
\begin{align}
\label{eqnSusyE}
\mQ_A^{\alpha}\mE_i&=\frac{1}{g}\left(\partial_t-ig\scal[[]{\mE_i}{\cdot}\right)X^{\alpha}_{iA}\\
\label{eqnSusyV}
\mQ_A^{\alpha}\mV_{i,i+1}&=iX^{\alpha}_{i+1\,A}\mV_{i,i+1}-i\mV_{i,i+1}X^{\alpha}_{iA}
\end{align}
\end{subequations}
Here, and in the following, we adopt the conventions of Ref.~\onlinecite{BDKM04}.

\section{Edge Operators}

The edge operators are computed as sketched in Ref.~\onlinecite{CH11}. One finds the following
solution of (\ref{eqnSusyE}), making use of the Euler-Lagrange equations.
\begin{align*}
\mE_i&=\frac{1}{2}\lambda_{i\beta}\tlambda_{i\dbeta}A^{\beta\dbeta}
+\frac{i}{c_0}\tlambda_{i\dbeta}\tpsi_A^{\dbeta}\eta_i^A
-\frac{i\sqrt{2}}{2c_0^2}\frac{\tlambda_{i\dbeta}\lambda_{(i-1)\gamma}D^{\dbeta\gamma}\ophi_{AB}}
{\scal{i}{i-1}}\eta_i^A\eta_i^B\\
&\qquad+\frac{1}{3c_0^3}\varepsilon_{ABCD}
\frac{\lambda_{(i-1)\xi}\tlambda_{i\dbeta}\lambda_{(i-1)\gamma}D^{\dbeta\gamma}
\psi^{\xi A}}{\scal{i}{i-1}^2}\eta_i^B\eta_i^C\eta_i^D\\
&\qquad+\frac{i}{24c_0^4}\varepsilon_{ABCD}
\frac{\lambda_{(i-1)\gamma}\lambda_{(i-1)\xi}\tlambda_{i\dbeta}\lambda_{(i-1)\beta}D^{\dbeta\beta}F^{\gamma\xi}}
{\scal{i}{i-1}^3}\eta_i^A\eta_i^B\eta_i^C\eta_i^D
\end{align*}
with
\begin{align*}
X^{\alpha}_{iA}:=\frac{g\lambda_{i-1}^{\alpha}}{c_0\scal{i}{i-1}}
&\left(-2i\sqrt{2}\,\ophi_{AB}\eta_i^B
+\varepsilon_{ABCD}\frac{2\lambda_{(i-1)\gamma}\psi^{\gamma B}}{c_0\scal{i}{i-1}}\eta_i^C\eta_i^D\right.\\
&\qquad\left.+\frac{i}{3c_0^2}\varepsilon_{ABCD}\frac{\lambda_{(i-1)\gamma}\lambda_{(i-1)\beta}F^{\gamma\beta}}{\scal{i}{i-1}^2}
\eta_i^B\eta_i^C\eta_i^D\right)
\end{align*}

\section{A Recursion Formula}

We expand
\begin{align}
\label{eqnVCoefficients}
\mV_{i,i+1}&=\sum_{k=0}^4\sum_{l=0}^4
V_{A_1\ldots A_k\;B_1\ldots B_l}\,\eta_i^{A_1}\ldots\eta_i^{A_k}\eta_{i+1}^{B_1}\ldots\eta_{i+1}^{B_l}
\end{align}
and, similarly, denote the coefficients of $X_{iA}^{\alpha}$ by
\begin{align*}
X_{iA}^{\alpha}=X_{iA}^{\alpha(1)}+X_{iA}^{\alpha(2)}+X_{iA}^{\alpha(3)}
=X_{iAA_1}^{\alpha(1)}\eta_i^{A_1}+X_{iAA_1A_2}^{\alpha(2)}\eta_i^{A_1}\eta_i^{A_2}
+X_{iAA_1A_2A_3}^{\alpha(3)}\eta_i^{A_1}\eta_i^{A_2}\eta_i^{A_3}
\end{align*}
Let $V_0=1$ (i.e. $\mV_{i,i+1}=1+\mO(\eta)$) and require that $\mV_{i\,i+1}$ only
depends on the generators $\eta_i$ and $\eta_{i+1}$. Then
(\ref{eqnSusyV}) with $X^{\alpha}_{iA}$ as above has the following
unique solution: All coefficients $V_{B_1,\ldots,B_d}=0$ for $d>0$ (i.e. all ''pure $\eta_{i+1}$-terms'') vanish and the remaining coefficients are successively determined by the following recursion formula.
\begin{align*}
&V_{A\,A_1\ldots A_k\,B_1\ldots B_l}\\
&\qquad=\frac{(-1)^{d+1}\lambda_{(i+1)\alpha}}{(k+1)c_0\scal{i+1}{i}}\left(-q_A^{\alpha}(V_{A_1\ldots A_k\,B_1\ldots B_l})
+iX_{(i+1)AB_l}^{\alpha(1)}V_{A_1\ldots A_k\,B_1\ldots B_{l-1}}\right.\\
&\qquad\qquad\qquad\left.+iX_{(i+1)AB_{l-1}B_l}^{\alpha(2)}V_{A_1\ldots A_k\,B_1\ldots B_{l-2}}
+iX_{(i+1)AB_{l-2}B_{l-1}B_l}^{\alpha(3)}V_{A_1\ldots A_k\,B_1\ldots B_{l-3}}\right.\\
&\qquad\qquad\qquad\left.-i(-1)^lV_{A_1\ldots A_{k-1}\,B_1\ldots B_l}X_{iAA_k}^{\alpha(1)}
-i(-1)^dV_{A_1\ldots A_{k-2}\,B_1\ldots B_l}X_{iAA_{k-1}A_k}^{\alpha(2)}\right.\\
&\qquad\qquad\qquad\left.-i(-1)^lV_{A_1\ldots A_{k-3}\,B_1\ldots B_l}X_{iAA_{k-2}A_{k-1}A_k}^{\alpha(3)}\right)
\end{align*}
where $d=k+l$.

\begin{proof}
For calculations, it is easier to work with an expansion where the
generators $\eta_i$ and $\eta_{i+1}$ can stand in any order:
\begin{align*}
\mV_{i,i+1}=\sum_{d=0}^8C_{B_1\ldots B_d}^{j_1\ldots j_d}\eta_{j_1}^{B_1}\ldots\eta_{j_d}^{B_d}\;,\quad
V_{A_1\ldots A_k\;B_1\ldots B_l}
=\left(\arr{c}{k+l\\k}\right)C^i_{A_1}{}^{\ldots}_{\ldots}{}^i_{A_k}{}^{i+1}_{B_1}{}_{\ldots}^{\ldots}{}^{i+1}_{B_l}
\end{align*}
with $j_i\in\{i,i+1\}$.
Now, applying from the left a fixed $\dd{\eta_k^A}$ in the $C$-expansion
kills the corresponding $\eta$ terms which can occur at every position, thus giving a
symmetry factor of $d$ and a sign such that
\begin{align*}
\mQ^{\alpha}_A(\mV_{i,i+1})
&=\sum_{d=0}^8\left(q_A^{\alpha}(C_{B_1\ldots B_d}^{j_1\ldots j_d})
+c_0(d+1)\,(-1)^{\abs{C_{AB_1\ldots B_d}^{kj_1\ldots j_d}}}
\lambda_k^{\alpha}\,C_{AB_1\ldots B_d}^{kj_1\ldots j_d}\right)\eta_{j_1}^{B_1}\ldots\eta_{j_d}^{B_d}
\end{align*}
(\ref{eqnSusyV}) is thus equivalent to the recursion formula
\begin{align*}
&c_0(d+1)\,(-1)^{\abs{C_{AB_1\ldots B_d}^{kj_1\ldots j_d}}}
\lambda_k^{\alpha}\,C_{AB_1\ldots B_d}^{kj_1\ldots j_d}\eta_{j_1}^{B_1}\ldots\eta_{j_d}^{B_d}\\
&\qquad=-q_A^{\alpha}(C_{B_1\ldots B_d}^{j_1\ldots j_d})\eta_{j_1}^{B_1}\ldots\eta_{j_d}^{B_d}
+i\sum_{k+l=d}({X^{\alpha}_{i+1\,A}}|_{\eta^k}{\mV_{i,i+1}}|_{\eta^l}-{\mV_{i,i+1}}|_{\eta^k}{X^{\alpha}_{iA}}|_{\eta^l})
\end{align*}

By induction, one shows that the coefficients are of parity $\abs{C_{B_1\ldots B_d}^{j_1\ldots j_d}}\equiv_2d$.
Also by induction, we see that all coefficients $C_{B_1\ldots B_d}^{i+1\ldots i+1}=0$ vanish:
In the recursion formula so far established, we consider the case $j_1=\ldots j_d=i+1$ and multiply both sides with $\lambda_{i\alpha}$.
Then only the left hand side with $k=i+1$ remains and
\begin{align*}
C_{AB_1\ldots B_d}^{i+1,i+1\ldots i+1}\eta_{i+1}^{B_1}\ldots\eta_{i+1}^{B_d}
=\frac{(-1)^{d+1}\lambda_{i\alpha}}{\scal{i}{i+1}c_0(d+1)}
\left(-q_A^{\alpha}(C_{B_1\ldots B_d}^{i+1\ldots i+1})\eta_{i+1}^{B_1}\ldots\eta_{i+1}^{B_d}\right)
\end{align*}
since $\lambda_{i\alpha}X^{\alpha}_{i+1\,A}=0$ and $X^{\alpha}_{i\,A}=\mO(\eta_i)$.
For $d=0$, the right hand side $\sim q_A^{\alpha}(1)=0$ vanishes and thus $C_B^{i+1}=0$.
Take this as induction basis and assume that $C_{B_1\ldots B_d}^{i+1\ldots i+1}=0$. The same
recursion formula then implies that $C_{AB_1\ldots B_d}^{i+1,i+1\ldots i+1}=0$.

Now, by multiplying both sides of the recursion formula with $\lambda_{(i+1)\alpha}$, only the left hand side
with $k=i$ remains, and we yield
\begin{align*}
&C_{AB_1\ldots B_d}^{ij_1\ldots j_d}\eta_{j_1}^{B_1}\ldots\eta_{j_d}^{B_d}\\
&\qquad=\frac{(-1)^{d+1}\lambda_{(i+1)\alpha}}{\scal{i+1}{i}c_0(d+1)}
\left(-q_A^{\alpha}(C_{B_1\ldots B_d}^{j_1\ldots j_d})\eta_{j_1}^{B_1}\ldots\eta_{j_d}^{B_d}
+(iX^{\alpha}_{i+1\,A}\mV_{i,i+1}-i\mV_{i,i+1}X^{\alpha}_{iA})|_{\eta^d}\right)
\end{align*}
Writing the second term on the right hand side in the $C$-expansion and then translating everything back
to the original expansion (\ref{eqnVCoefficients}) using
\begin{align*}
C_{AB_1\ldots B_d}^{ij_1\ldots j_d}\eta_{j_1}^{B_1}\ldots\eta_{j_d}^{B_d}|_{\eta_i^k\eta_{i+1}^l}
=\frac{k+1}{d+1}V_{A\,A_1\ldots A_k\,B_1\ldots B_l}\eta_i^{A_1}\ldots\eta_i^{A_k}\,\eta_{i+1}^{B_1}\ldots\eta_{i+1}^{B_l}
\end{align*}
the statement is finally obtained.
\end{proof}

\section{Vertex Operators}

By the recursion formula of the previous section, the coefficients of the vertex
operators in the expansion (\ref{eqnVCoefficients}) can be explicitly calculated.
Up to order three, the result reads
\begin{align*}
\mV_{i,i+1}&=1-\frac{\sqrt{2}\,gi_{\pm}}{c_0^2i_-i_+}\ophi_{A_1A_2}\,\eta_i^{A_1}\eta_i^{A_2}
+\frac{2\sqrt{2}\,g}{c_0^2i_+}\ophi_{A_1B_1}\,\eta_i^{A_1}\eta_{i+1}^{B_1}\\
&\qquad+\frac{2ig}{3c_0^3}\frac{i_{\pm}\left(-i_-\lambda_{(i+1)\gamma}+i_+\lambda_{(i-1)\gamma}\right)\psi^{\gamma C}}
{i_-^2i_+^2}\varepsilon_{A_1A_2A_3C}\,\eta_i^{A_1}\eta_i^{A_2}\eta_i^{A_3}\\
&\qquad+\frac{2ig\lambda_{(i+1)\gamma}\psi^{\gamma C}}{c_0^3i_+^2}\varepsilon_{A_1A_2B_1C}\,\eta_i^{A_1}\eta_i^{A_2}\eta_{i+1}^{B_1}
-\frac{2ig\lambda_{i\gamma}\psi^{\gamma C}}{c_0^3i_+^2}\varepsilon_{A_1B_1B_2C}\,\eta_i^{A_1}\eta_{i+1}^{B_1}\eta_{i+1}^{B_2}\\
&\qquad+\mO(\eta^4)
\end{align*}
with $i_-:=\scal{i}{i-1}$, $i_+:=\scal{i+1}{i}$ and $i_{\pm}:=\scal{i+1}{i-1}$.

\subsubsection*{Fourth Order}

The (non-vanishing) fourth order coefficients (\ref{eqnVCoefficients}) of $\mV_{i,i+1}$
are as follows.
\begin{align*}
V_{A_1A_2A_3A_4}
&=\left(\frac{gi_{\pm}\left(i_-^2\lambda_{(i+1)\beta}\lambda_{(i+1)\gamma}-i_-i_+\lambda_{(i-1)\beta}\lambda_{(i+1)\gamma}
+i_+^2\lambda_{(i-1)\beta}\lambda_{(i-1)\gamma}\right)F^{\beta\gamma}}{12c_0^4i_-^3i_+^3}\right.\\
&\qquad+\left.\frac{g^2i_{\pm}^2\ophi_{CD}\phi^{CD}}{12c_0^4i_-^2i_+^2}\right)\varepsilon_{A_1A_2A_3A_4}\\
V_{A_1A_2A_3B_1}
&=-\frac{g\lambda_{(i+1)\beta}\lambda_{(i+1)\gamma}F^{\beta\gamma}}{3c_0^4i_+^3}\varepsilon_{A_1A_2A_3B_1}
-\frac{4g^2i_{\pm}}{c_0^4i_-i_+^2}\ophi_{A_1B_1}\ophi_{A_2A_3}\\
V_{A_1A_2B_1B_2}
&=\frac{g\lambda_{i\beta}\lambda_{(i+1)\gamma}F^{\beta\gamma}}{2c_0^4i_+^3}\varepsilon_{A_1A_2B_1B_2}
-\frac{g^2}{c_0^4i_+^2}\scal[[]{\ophi_{A_1A_2}}{\ophi_{B_1B_2}}
-\frac{4g^2}{c_0^4i_+^2}\ophi_{A_1B_1}\ophi_{A_2B_2}\\
V_{A_1B_1B_2B_3}
&=-\frac{g\lambda_{i\beta}\lambda_{i\gamma}F^{\beta\gamma}}{3c_0^4i_+^3}\varepsilon_{A_1B_1B_2B_3}
\end{align*}

\subsubsection*{Fifth Order}

\begin{align*}
V_{A_1A_2A_3A_4B_1}
&=\frac{i\sqrt{2}\,g^2i_{\pm}}{3c_0^5i_-^2i_+^3}
\left(4(i_-\lambda_{(i+1)\gamma}-i_+\lambda_{(i-1)\gamma})
\varepsilon_{A_2A_3A_4C}\ophi_{A_1B_1}\psi^{\gamma C}\right.\\
&\qquad\qquad\qquad\left.
-6i_-\lambda_{(i+1)\gamma}\psi^{\gamma C}\varepsilon_{A_1A_2B_1C}\ophi_{A_3A_4}\right)\\
V_{A_1A_2A_3B_1B_2}
&=\frac{i\sqrt{2}\,g^2\lambda_{(i+1)\beta}}{3c_0^5i_+^3}\left(\varepsilon_{A_2A_3B_1B_2}\scal[[]{\ophi_{A_1C}}{\psi^{\beta C}}
+\varepsilon_{A_1A_2A_3C}\scal[[]{\ophi_{B_1B_2}}{\psi^{\beta C}}\right.\\
&\qquad\qquad\qquad\qquad-\varepsilon_{A_1B_1B_2C}\scal[[]{\ophi_{A_2A_3}}{\psi^{\beta C}}
-4\varepsilon_{A_1A_2B_1C}\psi^{\beta C}\ophi_{A_3B_2}\\
&\qquad\qquad\qquad\qquad\left.-8\varepsilon_{A_2A_3B_1C}\ophi_{A_1B_2}\psi^{\beta C}\right)\\
&\qquad+\frac{2i\sqrt{2}\,g^2i_{\pm}\lambda_{i\gamma}}{c_0^5i_-i_+^3}\varepsilon_{A_1B_1B_2C}\psi^{\gamma C}\ophi_{A_2A_3}\\
V_{A_1A_2B_1B_2B_3}
&=\frac{2i\sqrt{2}\,g^2\lambda_{i\gamma}}{3c_0^5i_+^3}
\left(-\scal[[]{\ophi_{A_1C}}{\psi^{\gamma C}}\varepsilon_{A_2B_1B_2B_3}
+3\scal[[]{\ophi_{A_1B_1}}{\psi^{\gamma C}}_+\varepsilon_{A_2B_2B_3C}\right)
\end{align*}
where $\scal[[]{X}{Y}_+:=XY+YX$ denotes the anticommutator.

\subsubsection*{Sixth Order}

\begin{align*}
V_{A_1A_2A_3A_4B_1B_2}
&=-\frac{\sqrt{2}g^2\lambda_{(i+1)\alpha}\lambda_{(i+1)\beta}}{24c_0^6i_+^4}
\varepsilon_{A_1A_2A_3A_4}\scal[[]{\ophi_{B_1B_2}}{F^{\beta\alpha}}\\
&\qquad+\frac{\sqrt{2}g^2\lambda_{(i+1)\alpha}\lambda_{(i+1)\beta}}{6c_0^6i_+^4}
\varepsilon_{A_1A_2A_3B_1}\left(F^{\beta\alpha}\ophi_{A_4B_2}+3\ophi_{A_4B_2}F^{\beta\alpha}\right)\\
&\qquad-\frac{\sqrt{2}g^2i_{\pm}}{2c_0^6i_-i_+^4}\lambda_{i\beta}\lambda_{(i+1)\gamma}F^{\beta\gamma}\ophi_{A_1A_4}\varepsilon_{A_2A_3B_1B_2}\\
&\qquad+\frac{\sqrt{2}\,g^3i_{\pm}}{2c_0^6i_-i_+^3}
\left(2\scal[[]{\ophi_{A_1A_2}}{\ophi_{B_1B_2}}\ophi_{A_3A_4}+8\ophi_{A_2B_1}\ophi_{A_3B_2}\ophi_{A_1A_4}\right)\\
&\qquad+\frac{g^2}{3c_0^6i_-^2i_+^4}\left(\varepsilon_{A_2A_3A_4C}\varepsilon_{A_1B_1B_2D}
(i_-^2\lambda_{(i+1)\gamma}\lambda_{(i+1)\delta})\right.\\
&\qquad\qquad\qquad+\varepsilon_{A_1B_1B_2C}\varepsilon_{A_2A_3A_4D}
(i_-^2\lambda_{(i+1)\gamma}\lambda_{(i+1)\delta}+4i_-i_{\pm}\lambda_{i\gamma}\lambda_{(i+1)\delta}\\
&\qquad\qquad\qquad\qquad\qquad\qquad\qquad\qquad-4i_+i_{\pm}\lambda_{i\gamma}\lambda_{(i-1)\delta})\\
&\qquad\qquad\qquad+\left.\varepsilon_{A_2A_3B_1C}\varepsilon_{A_1A_4B_2D}
(6i_-^2\lambda_{(i+1)\gamma}\lambda_{(i+1)\delta})\right)\psi^{\gamma C}\psi^{\delta D}
\end{align*}
and
\begin{align*}
V_{A_1A_2A_3B_1B_2B_3}
&=\frac{\sqrt{2}\,g^2\lambda_{i\gamma}\lambda_{(i+1)\alpha}}{9c_0^6i_+^4}
\left(\scal[[]{\ophi_{A_1A_2}}{F^{\gamma\alpha}}\varepsilon_{A_3B_1B_2B_3}\right.\\
&\qquad\qquad\qquad\left.+3\scal[[]{\ophi_{A_2B_1}}{F^{\gamma\alpha}}_+\varepsilon_{A_3B_2B_3A_1}
+3\ophi_{A_1B_3}F^{\gamma\alpha}\varepsilon_{A_2A_3B_1B_2}\right)\\
&\qquad+\frac{\sqrt{2}\,g^2i_{\pm}\lambda_{i\gamma}\lambda_{i\beta}F^{\gamma\beta}\ophi_{A_1A_2}}{3c_0^6i_-i_+^4}\varepsilon_{A_3B_1B_2B_3}\\
&\qquad+\frac{2\sqrt{2}\,g^3}{18c_0^6i_+^3}\left(\scal[[]{\ophi_{A_2C}}{\scal[[]{\ophi_{A_1D}}{\ophi_{EF}}}\varepsilon_{CDEF}\varepsilon_{A_3B_1B_2B_3}\right.\\
&\qquad\qquad\qquad+6\scal[[]{\ophi_{A_2B_1}}{\scal[[]{\ophi_{A_1A_3}}{\phi_{B_2B_3}}}_+\\
&\qquad\qquad\qquad\left.-6\ophi_{A_1B_3}\scal[[]{\ophi_{A_2A_3}}{\ophi_{B_1B_2}}
-24\ophi_{A_1B_3}\ophi_{A_2B_1}\ophi_{A_3B_2}\right)\\
&\qquad+\frac{4g^2\lambda_{i\gamma}\lambda_{(i+1)\alpha}}{9c_0^6i_+^4}
\left(-\varepsilon_{A_1A_2CD}\varepsilon_{A_3B_1B_2B_3}\scal[[]{\psi^{\alpha D}}{\psi^{\gamma C}}_+\right.\\
&\qquad\qquad\qquad+3\varepsilon_{A_3B_2B_3C}\varepsilon_{A_1A_2B_1D}\scal[[]{\psi^{\alpha D}}{\psi^{\gamma C}}\\
&\qquad\qquad\qquad\left.-3\varepsilon_{A_1B_2B_3C}\varepsilon_{A_2A_3B_1D}\psi^{\gamma C}\psi^{\alpha D}\right)\\
V_{A_1A_2B_1B_2B_3B_4}
&=\frac{\sqrt{2}\,g^2\lambda_{i\gamma}\lambda_{i\beta}\scal[[]{F^{\gamma\beta}}{\ophi_{A_1B_1}}_+\varepsilon_{A_2B_2B_3B_4}}{3c_0^6i_+^4}\\
&\qquad+\frac{2g^2\lambda_{i\gamma}\lambda_{i\delta}\psi^{\gamma C}\psi^{\delta D}}{c_0^6i_+^4}\varepsilon_{A_1B_3B_4C}\varepsilon_{A_2B_1B_2D}
\end{align*}

\subsubsection*{Seventh Order}

\begin{align*}
V_{A_1A_2A_3B_1B_2B_3B_4}
&=\frac{2ig^2\lambda_{i\gamma}\lambda_{i\beta}\lambda_{(i+1)\epsilon}}{9c_0^7i_+^5}
\varepsilon_{A_1A_2B_1C}\varepsilon_{A_3B_2B_3B_4}\left(2F^{\gamma\beta}\psi^{\epsilon C}+\psi^{\epsilon C}F^{\gamma\beta}\right)\\
&\qquad-\frac{ig^2\lambda_{i\epsilon}\lambda_{i\beta}\lambda_{(i+1)\gamma}}{3c_0^7i_+^5}
\varepsilon_{A_1B_1B_2C}\varepsilon_{A_2A_3B_3B_4}\left(F^{\gamma\beta}\psi^{\epsilon C}+2\psi^{\epsilon C}F^{\gamma\beta}\right)\\
&\qquad+\frac{8ig^3\lambda_{i\gamma}\scal[[]{\scal[[]{\ophi_{A_1C}}{\psi^{\gamma C}}}{\ophi_{A_2B_1}}_+\varepsilon_{A_3B_2B_3B_4}}{9c_0^7i_+^4}\\
&\qquad+\frac{2ig^3\lambda_{i\gamma}}{3c_0^7i_+^4}\varepsilon_{A_3B_3B_4D}
\scal[[]{\scal[[]{\ophi_{A_1A_2}}{\ophi_{B_1B_2}}}{\psi^{\gamma D}}_+\\
&\qquad+\frac{8ig^3\lambda_{i\gamma}}{9c_0^7i_+^4}\ophi_{A_1B_4}
\left(\scal[[]{\ophi_{A_2C}}{\psi^{\gamma C}}\varepsilon_{A_3B_1B_2B_3}
-3\scal[[]{\ophi_{A_2B_1}}{\psi^{\gamma C}}_+\varepsilon_{A_3B_2B_3C}\right)\\
&\qquad+\frac{2ig^3\lambda_{i\gamma}}{3c_0^7i_+^4}\varepsilon_{A_1B_3B_4C}\psi^{\gamma C}\scal[[]{\ophi_{A_2A_3}}{\ophi_{B_1B_2}}\\
&\qquad+\frac{8ig^3\lambda_{i\gamma}}{3c_0^7i_+^4}\varepsilon_{A_1B_3B_4C}\psi^{\gamma C}\ophi_{A_2B_1}\ophi_{A_3B_2}
\end{align*}
and
\begin{align*}
V_{A_1A_2A_3A_4B_1B_2B_3}&=V_{A_1A_2A_3A_4B_1B_2B_3}|_{\phi\phi\psi}+V_{A_1A_2A_3A_4B_1B_2B_3}|_{F\psi}
\end{align*}
with
\begin{align*}
&V_{A_1A_2A_3A_4B_1B_2B_3}|_{\phi\phi\psi}\\
&\qquad=\frac{ig^3\lambda_{(i+1)\beta}}{9c_0^7i_+^4}
\left(4\varepsilon_{A_4B_1B_2B_3}\scal[[]{\ophi_{A_1A_2}}{\scal[[]{\ophi_{A_3C}}{\psi^{\beta C}}}-2\varepsilon_{A_3A_4B_2B_3}\scal[[]{\ophi_{A_1B_1}}{\scal[[]{\ophi_{A_2C}}{\psi^{\beta C}}}_+\right.\\
&\qquad\qquad\qquad\qquad
-\varepsilon_{A_4B_1B_2B_3}\scal[[]{\psi^{\beta C}}{\scal[[]{\ophi_{A_1A_2}}{\ophi_{A_3C}}}
-2\varepsilon_{A_4B_1B_2B_3}\scal[[]{\ophi_{A_3C}}{\scal[[]{\psi^{\beta C}}{\ophi_{A_1A_2}}}\\
&\qquad\qquad\qquad\qquad
+3\varepsilon_{A_1A_2A_4C}\scal[[]{\ophi_{A_3B_1}}{\scal[[]{\psi^{\beta C}}{\phi_{B_2B_3}}}_++3\varepsilon_{A_1B_2B_3C}\scal[[]{\ophi_{A_3B_1}}{\scal[[]{\ophi_{A_2A_4}}{\psi^{\beta C}}}_+\\
&\qquad\qquad\qquad\qquad-\varepsilon_{A_4B_1B_2B_3}\scal[[]{\psi^{\beta C}}{\scal[[]{\ophi_{A_1A_2}}{\ophi_{A_3C}}}
-9\varepsilon_{A_2A_3B_1C}\scal[[]{\psi^{\beta C}}{\scal[[]{\ophi_{A_1A_4}}{\ophi_{B_2B_3}}}_+\\
&\qquad\qquad\qquad\qquad-12\varepsilon_{A_1A_2B_3C}\psi^{\beta C}\ophi_{A_3B_1}\ophi_{A_4B_2}\\
&\qquad\qquad\qquad\qquad+\ophi_{A_1B_3}\left(6\varepsilon_{A_2A_3A_4C}\scal[[]{\psi^{\beta C}}{\ophi_{B_1B_2}}
+6\varepsilon_{A_2B_1B_2C}\scal[[]{\ophi_{A_3A_4}}{\psi^{\beta C}}\right.\\
&\qquad\qquad\qquad\qquad\qquad\qquad+24\varepsilon_{A_2A_3B_1C}\scal[[]{\psi^{\beta C}}{\ophi_{A_4B_2}}_++12\varepsilon_{A_3A_4B_1C}\ophi_{A_2B_2}\psi^{\beta C}\\
&\qquad\qquad\qquad\qquad\qquad\qquad\left.\left.-5\varepsilon_{A_3A_4B_1B_2}\scal[[]{\ophi_{A_2C}}{\psi^{\beta C}}\right)\right)\\
&\qquad\qquad+\frac{4ig^3i_{\pm}\lambda_{i\gamma}}{3c_0^7i_-i_+^4}
\left(\varepsilon_{A_4B_1B_2B_3}\scal[[]{\ophi_{A_1C}}{\psi^{\gamma C}}\ophi_{A_2A_3}
-3\varepsilon_{A_2B_1B_2C}\scal[[]{\ophi_{A_1B_3}}{\psi^{\gamma C}}_+\ophi_{A_3A_4}\right)
\end{align*}
and
\begin{align*}
&V_{A_1A_2A_3A_4B_1B_2B_3}|_{F\psi}\\
&\qquad=\frac{ig^2\lambda_{(i+1)\alpha}\lambda_{i\gamma}\lambda_{(i+1)\beta}}{18c_0^7i_+^5}\left(9\varepsilon_{A_1A_2B_1C}\varepsilon_{A_3A_4B_2B_3}\scal[[]{\psi^{\beta C}}{F^{\gamma\alpha}}_++6\varepsilon_{A_2B_2B_3C}\varepsilon_{A_3A_4B_1A_1}\psi^{\gamma C}F^{\beta\alpha}\right.\\
&\qquad\qquad\qquad\qquad\left.+\varepsilon_{A_1A_2A_3C}\varepsilon_{A_4B_1B_2B_3}\scal[[]{F^{\beta\alpha}}{\psi^{\gamma C}}
+3\varepsilon_{A_4B_2B_3C}\varepsilon_{A_2A_3B_1A_1}\scal[[]{F^{\beta\alpha}}{\psi^{\gamma C}}_+\right)\\
&\qquad\qquad+\frac{2ig^2i_{\pm}\lambda_{(i+1)\delta}\lambda_{i\gamma}\lambda_{i\beta}}{9c_0^7i_-i_+^5}\varepsilon_{A_1A_2A_3C}\varepsilon_{A_4B_1B_2B_3}F^{\gamma\beta}\psi^{\delta C}\\
&\qquad\qquad+\frac{2ig^2i_{\pm}\lambda_{i\beta}\lambda_{i\gamma}\lambda_{(i-1)\delta}}{9c_0^7i_-^2i_+^4}
\varepsilon_{A_1B_1B_2B_3}\varepsilon_{A_2A_3A_4C}F^{\beta\gamma}\psi^{\delta C}
\end{align*}

\subsubsection*{Eighth Order}

\begin{align*}
V_{A_1A_2A_3A_4B_1B_2B_3B_4}&=V_{A_1A_2A_3A_4B_1B_2B_3B_4}|_{\phi^4}+V_{A_1A_2A_3A_4B_1B_2B_3B_4}|_{F\phi\phi}\\
&\qquad+V_{A_1A_2A_3A_4B_1B_2B_3B_4}|_{\phi\psi\psi}+V_{A_1A_2A_3A_4B_1B_2B_3B_4}|_{FF}
\end{align*}
with
\begin{align*}
V_{A_1A_2A_3A_4B_1B_2B_3B_4}|_{\phi^4}
&=\frac{g^4}{18c_0^8i_+^4}
\left(-2\varepsilon_{DEFG}\varepsilon_{A_4B_2B_3B_4}\scal[[]{\scal[[]{\ophi_{A_2D}}{\scal[[]{\ophi_{A_1E}}{\ophi_{FG}}}_-}{\ophi_{A_3B_1}}_+\right.\\
&\qquad\qquad\qquad+3\scal[[]{\scal[[]{\ophi_{A_2A_3}}{\ophi_{B_1B_2}}}{\scal[[]{\ophi_{A_1A_4}}{\ophi_{B_3B_4}}}_+\\
&\qquad\qquad\qquad-2\ophi_{A_2B_4}\scal[[]{\ophi_{A_3D}}{\scal[[]{\ophi_{A_1E}}{\ophi_{FG}}}\varepsilon_{A_4B_1B_2B_3}\varepsilon_{DEFG}\\
&\qquad\qquad\qquad-12\ophi_{A_2B_4}\scal[[]{\ophi_{A_3B_1}}{\scal[[]{\ophi_{A_1A_4}}{\ophi_{B_2B_3}}}_+\\
&\qquad\qquad\qquad+3\scal[[]{\ophi_{A_1A_2}}{\ophi_{B_3B_4}}\scal[[]{\ophi_{A_3A_4}}{\ophi_{B_1B_2}}\\
&\qquad\qquad\qquad+12\scal[[]{\ophi_{A_1A_2}}{\ophi_{B_3B_4}}\ophi_{A_3B_1}\ophi_{A_4B_2}\\
&\qquad\qquad\qquad+2\ophi_{A_1B_4}\scal[[]{\ophi_{A_3C}}{\scal[[]{\ophi_{A_2D}}{\ophi_{EF}}}\varepsilon_{CDEF}\varepsilon_{A_4B_1B_2B_3}\\
&\qquad\qquad\qquad+12\ophi_{A_1B_4}\scal[[]{\ophi_{A_3B_1}}{\scal[[]{\ophi_{A_2A_4}}{\phi_{B_2B_3}}}_+\\
&\qquad\qquad\qquad-12\ophi_{A_1B_4}\ophi_{A_2B_3}\scal[[]{\ophi_{A_3A_4}}{\ophi_{B_1B_2}}\\
&\qquad\qquad\qquad\left.-48\ophi_{A_1B_4}\ophi_{A_2B_3}\ophi_{A_3B_1}\ophi_{A_4B_2}\right)
\end{align*}
and
\begin{align*}
&V_{A_1A_2A_3A_4B_1B_2B_3B_4}|_{F\phi\phi}\\
&\qquad=\frac{g^3\lambda_{i\beta}\lambda_{(i+1)\gamma}}{36c_0^8i_+^5}
\left(-9\varepsilon_{A_3A_4B_3B_4}\scal[[]{F^{\gamma\beta}}{\scal[[]{\ophi_{A_1A_2}}{\ophi_{B_1B_2}}}_+\right.\\
&\qquad\qquad\qquad\qquad+4\varepsilon_{A_4B_2B_3B_4}\scal[[]{\scal[[]{\ophi_{A_2A_1}}{F^{\gamma\beta}}_-}{\ophi_{A_3B_1}}_+\\
&\qquad\qquad\qquad\qquad+8\ophi_{A_1B_4}\left(\scal[[]{\ophi_{A_2A_3}}{F^{\gamma\beta}}\varepsilon_{A_4B_1B_2B_3}
+3\scal[[]{\ophi_{A_2B_1}}{F^{\gamma\beta}}_+\varepsilon_{A_3B_2B_3A_4}\right)\\
&\qquad\qquad\qquad\qquad\left.
-12\varepsilon_{A_1A_2B_1B_2}\scal[[]{F^{\gamma\beta}}{\ophi_{A_3B_3}\ophi_{A_4B_4}}_+\right)\\
&\qquad\qquad-\frac{2g^3i_{\pm}\lambda_{i\gamma}\lambda_{i\beta}}{3c_0^8i_-i_+^5}
\varepsilon_{A_1B_1B_2B_3}\scal[[]{F^{\gamma\beta}}{\ophi_{A_2B_4}}_+\ophi_{A_3A_4}
\end{align*}
and
\begin{align*}
&V_{A_1A_2A_3A_4B_1B_2B_3B_4}|_{\phi\psi\psi}\\
&\qquad=\frac{\sqrt{2}\,g^3\lambda_{i\gamma}\lambda_{(i+1)\beta}}{18c_0^8i_+^5}
\left(8\varepsilon_{A_2A_3B_1D}\varepsilon_{A_4B_2B_3B_4}
\scal[[]{\scal[[]{\ophi_{A_1C}}{\psi^{\gamma C}}}{\psi^{\beta D}}_-\right.\\
&\qquad\qquad\qquad\qquad-12\varepsilon_{A_2A_3B_4D}
\varepsilon_{A_4B_2B_3C}\psi^{\beta D}\scal[[]{\ophi_{A_1B_1}}{\psi^{\gamma C}}_+\\
&\qquad\qquad\qquad\qquad-3\varepsilon_{A_2B_1B_2C}\varepsilon_{A_3A_4B_3B_4}
\left(\scal[[]{\ophi_{A_1D}}{\psi^{\beta D}}\psi^{\gamma C}
-2\psi^{\gamma C}\scal[[]{\ophi_{A_1D}}{\psi^{\beta D}}\right)\\
&\qquad\qquad\qquad\qquad+4\varepsilon_{A_4B_2B_3B_4}\left(\varepsilon_{A_1A_2CD}\scal[[]{\scal[[]{\psi^{\beta D}}{\psi^{\gamma C}}_+}{\ophi_{A_3B_1}}_+\right.\\
&\qquad\qquad\qquad\qquad\qquad\qquad\qquad\left.+\varepsilon_{A_1A_3B_1D}\scal[[]{\psi^{\beta D}}{\scal[[]{\ophi_{A_2C}}{\psi^{\gamma C}}}_-\right)\\
&\qquad\qquad\qquad\qquad+3\varepsilon_{A_4B_3B_4C}\left(\varepsilon_{A_1A_2A_3D}\scal[[]{\scal[[]{\psi^{\beta D}}{\ophi_{B_1B_2}}}{\psi^{\gamma C}}_-\right.\\
&\qquad\qquad\qquad\qquad\qquad\qquad\qquad\left.+\varepsilon_{A_1B_1B_2D}\scal[[]{\scal[[]{\ophi_{A_2A_3}}{\psi^{\beta D}}}{\psi^{\gamma C}}_-\right)\\
&\qquad\qquad\qquad\qquad+4\ophi_{A_1B_4}\left(-2\varepsilon_{A_2A_3CD}\varepsilon_{A_4B_1B_2B_3}\scal[[]{\psi^{\beta D}}{\psi^{\gamma C}}_+\right.\\
&\qquad\qquad\qquad\qquad\qquad\qquad\qquad\left.+6\varepsilon_{A_4B_2B_3C}\varepsilon_{A_2A_3B_1D}\scal[[]{\psi^{\beta D}}{\psi^{\gamma C}}_-\right.\\
&\qquad\qquad\qquad\qquad\qquad\qquad\qquad\left.-3\varepsilon_{A_2B_2B_3C}\varepsilon_{A_3A_4B_1D}\psi^{\gamma C}\psi^{\beta D}\right)\\
&\qquad\qquad\qquad\qquad+3\varepsilon_{A_1B_3B_4C}\psi^{\gamma C}
\left(2\varepsilon_{A_2A_3A_4D}\scal[[]{\psi^{\beta D}}{\ophi_{B_1B_2}}
+2\varepsilon_{A_2B_1B_2D}\scal[[]{\ophi_{A_3A_4}}{\psi^{\beta D}}\right.\\
&\qquad\qquad\qquad\qquad\qquad\qquad\qquad
-\varepsilon_{A_2A_3B_1B_2}\scal[[]{\ophi_{A_4D}}{\psi^{\beta D}}
+8\varepsilon_{A_2A_3B_1D}\scal[[]{\ophi_{A_4B_2}}{\psi^{\beta D}}_+\\
&\qquad\qquad\qquad\qquad\qquad\qquad\qquad\left.\left.+4\varepsilon_{A_2A_3B_1D}\ophi_{A_4B_2}\psi^{\beta D}\right)\right)\\
&\qquad\qquad-\frac{2\sqrt{2}g^3i_{\pm}\lambda_{i\beta}\lambda_{i\gamma}}{c_0^8i_-i_+^5}\varepsilon_{A_1B_3B_4C}\varepsilon_{A_2B_1B_2D}\psi^{\gamma C}\psi^{\beta D}\ophi_{A_3A_4}
\end{align*}
and
\begin{align*}
V_{A_1A_2A_3A_4B_1B_2B_3B_4}|_{FF}
&=-\frac{g^2\lambda_{i\gamma}\lambda_{i\beta}\lambda_{(i+1)\epsilon}\lambda_{(i+1)\alpha}}{36c_0^8i_+^6}\varepsilon_{A_1A_2A_3B_1}\varepsilon_{A_4B_2B_3B_4}
\left(3F^{\gamma\beta}F^{\epsilon\alpha}+F^{\epsilon\alpha}F^{\gamma\beta}\right)\\
&\qquad+\frac{g^2\lambda_{i\epsilon}\lambda_{i\beta}\lambda_{(i+1)\gamma}\lambda_{(i+1)\alpha}}{8c_0^8i_+^6}\varepsilon_{A_1A_2B_1B_2}\varepsilon_{A_3A_4B_3B_4}F^{\gamma\beta}F^{\epsilon\alpha}
\end{align*}

\subsubsection*{The General Structure}

From the above formulas for $\mV_{i\,i+1}$, we see that higher order terms factor into terms
with the structure of lower order terms:
\begin{align*}
\mV_{i,i+1}\sim\sum\prod
\left(1+\ophi\cdot\eta^2+\frac{1}{\sqrt{g}}\psi\cdot\eta^3+\frac{1}{g}F\cdot\eta^4\right)
\end{align*}
It is understood that this is not an equation but only a similarity which helps memorise the types of terms occurring.

\section*{Acknowledgements}

I would like to thank Jan Plefka, Johannes Henn and Konstantin Wiegandt for
interesting discussions.

\end{document}